\documentclass[proceedings]{rmaa}
\usepackage{rmaacite}
\usepackage{epsf}

\renewcommand{\P}[1]{%
\ifnum#1=1\hbox{OW~168--326E}\fi
\ifnum#1=2\hbox{OW~167--317}\fi
\ifnum#1=3\hbox{OW~163--317}\fi
\ifnum#1=5\hbox{OW~158--323}\fi
\ifnum#1=0\hbox{OW~171--334}\fi}

\title{The Effect of Gasdynamics on the Structure of Dark Matter Halos}

\author{Marcelo Alvarez\altaffilmark{1,2}, Paul R. 
Shapiro\altaffilmark{1}, and Hugo Martel\altaffilmark{1}
\affil{University of Texas at Austin}}

\altaffiltext{1}{Department of Astronomy.}
\altaffiltext{2}{Center for Computational Visualization, Texas Institute
for Computational and Applied Mathematics.}

\nonstopmode

\fulladdresses{
\item Marcelo Alvarez, Paul Shapiro, Hugo Martel: Department of Astronomy,
University of Texas, Austin, TX 78712, USA
(marcelo,shapiro,hugo@astro.as.utexas.edu)}

\shortauthor{Alvarez, Shapiro, \& Martel}
\shorttitle{The Effect of Gasdynamics on Dark Matter Halos}


\abstract{  

Adaptive SPH and N-body simulations were carried out to study the effect
of gasdynamics on the structure of dark matter halos that result from the
gravitational instability and fragmentation of cosmological pancakes. Such
halos resemble those formed in a hierarchically clustering CDM universe
and serve as a test-bed model for studying halo dynamics.  With no gas,
the density profile is close to the universal profile identified
previously from N-body simulations of structure formation in CDM. When gas
is included, the gas
in the halo is approximately isothermal, and both the dark matter and the
gas have singular central density profiles which are steeper than that of
the dark matter with no gas.  This worsens the disagreement between
observations of constant density cores in cosmological halos and the
singular ones found in simulations. We also find that the dark matter
velocity distribution is less isotropic than found by N-body simulations
of CDM, because of the strongly filamentary
substructure.}

\listofauthors{M. Alvarez, P.R. Shapiro, \& H. Martel}
\indexauthor{Alvarez, M.}
\indexauthor{Shapiro, P.R.}
\indexauthor{Martel, H.}

\nonstopmode

\begin{document}
\maketitle

\section{Introduction}
\label{sec:intro}

N-body simulations of structure formation that have as initial conditions
a spectrum of density fluctuations consistent with a cold dark matter
(CDM) universe have revealed that dark matter halos possess a universal
density profile that diverges as $r^{-\gamma}$ near the center, with
$1\leq\gamma\leq2$ (e.g. Navarro, Frenk, \& White 1997 ``NFW"; Moore
{\em{et al.}} 1998). Much effort has been made attempting to understand
the apparent discrepancy between these singular density profiles found in
N-body simulations and current observations of the rotation curves of
nearby dwarf galaxies and of strong lensing of background galaxies by the
galaxy cluster CL0024+1654 \cite{Tyson98}, which suggest that cosmological
halos of all scales have flat density cores, instead.  Several ways to
reconcile the theory with observations have been hypothesized, among which
are:

\begin{itemize}
\item 
Problems with the N-body gravity solvers.  Upon further scrutiny, however,
the simulations have been found to represent faithfully the actual
gravitational processes at work in hierarchical clustering.
\item 
Self-interacting dark matter.  If the dark matter is not collisionless as
previously assumed but rather can interact nongravitationally with itself,
this may change the dynamics in such a way as to produce constant density
cores
\cite{SS2000}.
\item
Gasdynamical effects due to the baryonic component.  The inclusion of gas
dynamics in the N-body simulations may lead to constant density cores.

\end{itemize}
It is the latter case which we address here.  In particular, we analyze
the density profiles of cosmological halos formed in numerical simulations
of cosmological pancakes with and without a substantial baryonic
component.

\section{Halo Formation by Gravitational Instability During Pancake
Collapse}

\label{sec:pancake}

Consider the growing mode of a single sinusoidal plane-wave density
fluctuation of comoving wavelength $\lambda_p$ and dimensionless
wavevector ${\bf
k_p=\hat{x}}$ (length unit = $\lambda_p$) in an Einstein-de Sitter
universe
dominated by cold, collisionless dark matter. Let the initial amplitude
$\delta_i$ at scale factor $a_i$ be chosen so that a density caustic forms 
in the collisionless component at scale factor $a=a_c=a_i/\delta_i$.

\begin{figure}
  \begin{center}
    \leavevmode
    {\rotatebox{90}{\includegraphics[width=2.9in]{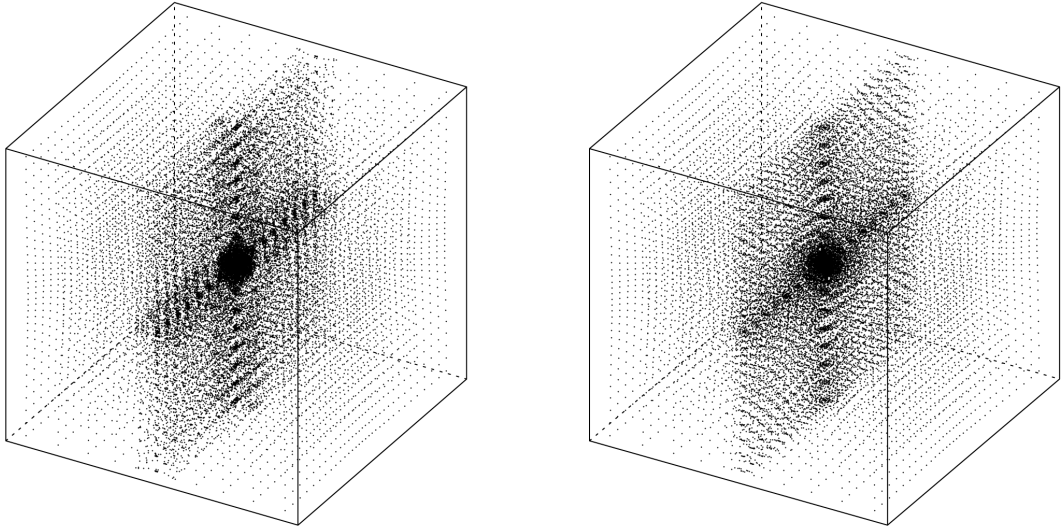}}}
    \caption{Dark matter (left) and Gas (right) particle positions at
    $a/a_c=3$ (volume $=\lambda_p^3$).}
    \label{fig:part}
  \end{center}
\end{figure}

Pancakes modelled in this way have been shown to be gravitationally
unstable, leading to filamentation and fragmentation during the collapse
\cite{Val97}.  Such instability can be triggered if we perturb the 1D
fluctuation described above by adding to the initial conditions two
transverse, plane-wave density fluctuations with equal wavelength
$\lambda_s=\lambda_p$, wavevectors ${\bf k_s}$ pointing along the
orthogonal vectors ${\bf \hat{y}}$ and ${\bf \hat{z}}$, and amplitudes
$\epsilon_y\delta_i$ and $\epsilon_z\delta_i$, respectively, where
$\epsilon_y \ll 1$ and $\epsilon_z \ll 1$.  A pancake perturbed by two
such density perturbations will be referred to as
$S_{1,\epsilon_y,\epsilon_z}$.  All results presented here refer to the
case $S_{1,0.2,0.2}$ unless otherwise noted.  Such a perturbation leads to
the formation of a quasi-spherical mass concentration in the pancake plane
at the intersection of two filaments (Figure 1).  Halos formed from
pancake collapse as modelled above have a density profile similar in shape
to those found in simulations of hierarchical structure formation with
realistic initial fluctuation spectra \cite{Val96,Mar00}.  As such,
pancake collapse and fragmentation can be used as a test-bed model for
halo formation which retains the realistic features of anisotropic
collapse, continuous infall, and cosmological boundary conditions.  
Questions we seek to answer quantitatively are:

\begin{itemize}

\item Is the halo that results from pancake
instability and fragmentation in a relaxed equilibrium state?
\item How isotropic is the resulting velocity
distribution for the dark matter, and how isothermal are the gas and dark 
matter?
\item What effect does including gas in the simulations have on
the dark matter density profile?

\end{itemize} \begin{figure}
  \begin{center}
    \leavevmode
    \includegraphics[width=5.3in]{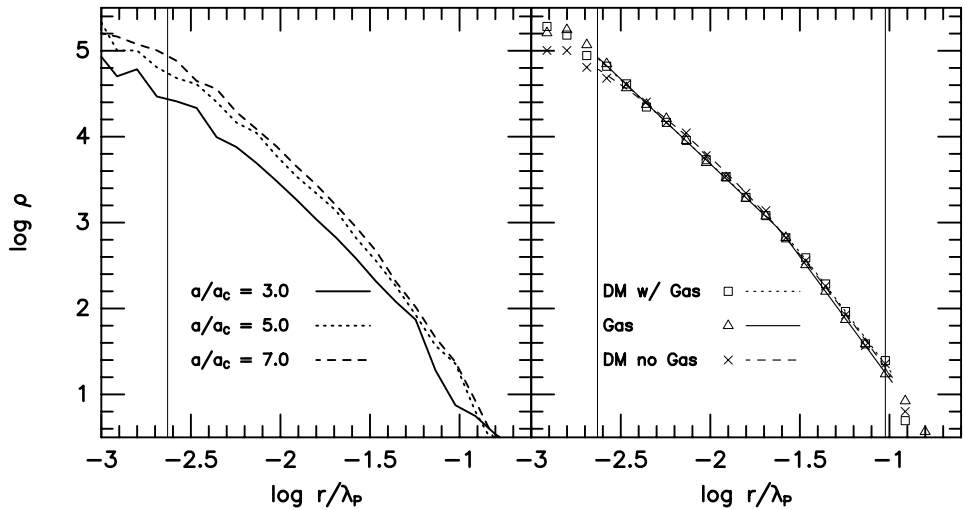}
    \caption{(left) Evolution of dark matter density profile found by
spherically averaging over logarithmically-spaced radial bins.  (right) 
Density profiles for each component (in units of cosmic mean density
for that component) at $a/a_c=5.0$. Label ``DM w/ Gas"
means the dark matter profile in the simulation which includes gas.}
    \label{fig:part}
  \end{center}
\end{figure}

The pancake problem (without radiative cooling) is self-similar and
scale-free, once distance is
expressed in units of the pancake wavelength $\lambda_p$ and time is
expressed in terms of the cosmic scale factor $a$ in units of the scale factor $a_c$ at which caustics form in the dark matter and shocks in
the gas.  In the currently-favored flat, cosmological-constant-dominated
universe, however, this self-similarity is broken because
$\Omega_M/\Omega_\Lambda$ decreases with time, where $\Omega_M$ and
$\Omega_\Lambda$ are the matter and vacuum energy density parameters,
respectively. For objects which
collapse at high redshift in such a universe (e.g. dwarf galaxies), the
Einstein-de Sitter results are still applicable as long as we take
$(\Omega_B/\Omega_{DM})_{E-deS}=(\Omega_B/\Omega_{DM})_{\Lambda}$, where
$\Omega_B$ and $\Omega_{DM}$ are the baryon and dark matter density
parameters.  If $\Omega_B=0.045$, $\Omega_{DM}=0.255$,
and $\Omega_\Lambda=0.7$ at present, then the E-deS results are applicable
if we take $\Omega_B=0.15$ and $\Omega_{DM}=0.85$, instead.

\section{Results}
\label{sec:results}

\begin{figure}
  \begin{center}
    \leavevmode
    \includegraphics[width=6in]{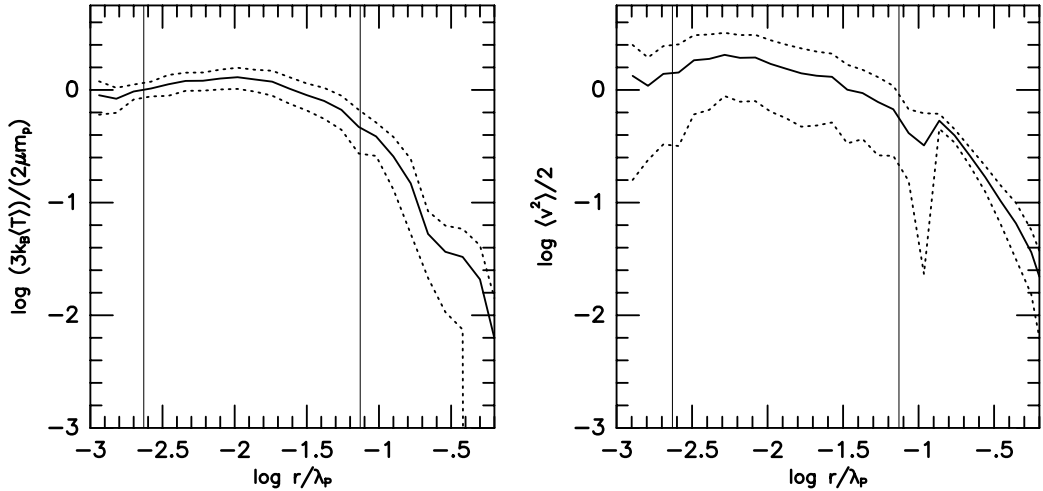}
    \caption{At $a/a_c=5.0$, Gas (left) and Dark Matter (right)
specific thermal
and kinetic energy profiles, respectively, in dimensionless computational
units, at $a/a_c=5.0$, which resulted from spherically averaging
logarithmically-spaced radial bins. The dotted lines show the rms
fluctuations. The two vertical lines in each denote the numerical
softening length for the gravity calculation and $r_{200}$.}
    \label{fig:spread}
  \end{center}
\end{figure}

The numerical technique used here to simulate the collapse and
fragmentation of a cosmological pancake couples the gasdynamical method
Adaptive Smoothed Particle Hydrodynamics (ASPH) \cite{Sha96,Owen98} with a
particle-particle, particle-mesh (P$^3$M) N-body gravity solver
\cite{Mar200}. Results presented here are for a simulation with $64^3$
particles each of dark matter and gas when hydrodynamics is included, and
$64^3$ dark matter particles for the case with gravity only, on a 
P$^3$M grid with $128^3$ cells. 
We consider an Einstein-de Sitter universe
($\Omega_M=1$, $\Omega_\Lambda=0$) and two cases:
$(\Omega_B,\Omega_{DM})=(0,1)$ or $(0.15,0.85)$.  The
pancake-filament-halo structure of our results is illustrated by Figure 1.
By $a/a_c=5$, this halo contains approximately $10^5$ particles each of
dark matter and gas.

\subsection{The equilibrium halo structure and similarity to CDM $N$-body
simulation halos}

The left panel of Figure 2 shows the evolution of the density profile of
the dark matter halo in the simulation with no gas.  By $a/a_c=5$, the
halo has asymptoted to an equilibrium profile.  A $\chi^2$-minimization
fit was performed on this profile over the range between the gravitational
softening length $r_{soft}$ and $r_{200}$, the radius within which the
mean density is 200 times the cosmic mean density at that epoch, where
$r_{soft}/\lambda_p\cong2.3\times10^{-3}$ and
$r_{200}/\lambda_p\cong7.5\times10^{-2}$, for each component X of the
matter density. The fitting function we use, in general, is of the form

\begin{equation}
\rho_X=\frac{\rho_s}{{\big{(}}\frac{r}{r_s})^{\gamma}(1+(\frac{r}{r_s})^{{1}/{\alpha}
})^{\alpha(\beta-\gamma)}} 
\end{equation} 
\cite{Zhao96,Hern90}, which approaches $\rho_X\ {\propto}\ r^{-\gamma}$ at
$r\ {\ll}\ r_s$ and $\rho_X\ {\propto}\ r^{-\beta}$ at $r\ {\gg}\ r_s$,
where $\rho_X$ and $\rho_s$ are in units of the mean cosmic density of
component X at the same epoch. We found
$\lbrace\alpha,\beta,\gamma,r_s/\lambda_p,\rho_s\rbrace=
\lbrace0.866,3.18,1.18,1.39\times10^{-2},9.19\times10^3\rbrace$.  This
profile is similar in shape to the universal density profile found for CDM
halos by NFW based upon N-body simulations, for which
$\lbrace\alpha,\beta,\gamma\rbrace=\lbrace1,3,1\rbrace$.  In the notation
of NFW, our best-fit halo concentration parameter, $c{\equiv}r_{200}/r_s$,
is $c\cong5$, while our $\rho_s$ corresponds to NFW's $\delta_c$.  
According to NFW, $\delta_c=(200/3)c^3/[ln(1+c)-c/(1+c)]$.  For our value
of $c\cong5$, this yields $\delta_c\cong9\times10^{3}$, very close to our
best-fit value of $\rho_s$. Such a density profile is thus a more general
outcome of cosmological halo formation by gravitational condensation, not
limited to hierarchical clustering.

\begin{figure}
  \begin{center}
    \leavevmode
    \includegraphics[width=4.in]{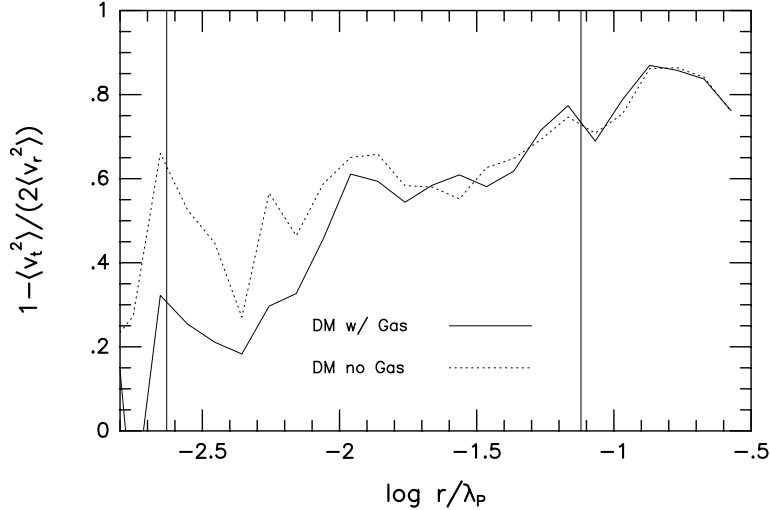}
    \caption{The anisotropy parameter $\beta_{an}$ for the dark matter in
the cases with and without gas, calculated by spherically averaging the
tangential and radial components within logarithmically-spaced
radial bins.}
    \label{fig:part}
  \end{center}
\end{figure}

A common assumption is that ``virialized" objects resulting from
cosmological perturbations have an approximately isothermal temperature
distribution and a dark matter velocity distribution that is isotropic.
Figure 3 shows the spherically-averaged specific thermal energy of the gas
($\frac{3}{2}k_B{\langle}T{\rangle}/{\mu}m_p$) and kinetic energy of the
dark matter ($\frac{1}{2}{\langle}v^2{\rangle}$). The gas in the halo is
nearly isothermal from the softening length to $r_{200}$, with $\langle
T(r_{soft}) \rangle /\langle T(r_{200}) \rangle \leq 2$.  The dark matter
kinetic energy follows a similar profile with a slightly steeper drop
closer to the center. Figure 4 shows the parameter $\beta_{an} \equiv
1-{\langle}v_t^2{\rangle}/2{\langle}v_r^2{\rangle}$, which is a measure of
the anisotropy of the velocity distribution of the collisionless dark
matter, as a function of distance from the center.  In the limits of
isotropic and radial motion, $\beta_{an}=0$ and $1$, respectively.  The
distribution is somewhat more anisotropic than usually found in
hierarchical clustering simulations, due to the highly anisotropic infall
which channels matter along the filaments that intersect at our halo,
favoring radial motion.  While this is true for both cases, the case with
gas included seems to suppress radial motion at small radii.

\subsection{Comparison of density profiles with and without gas}

Figure 2 shows the profile fits for each component.  The parameters
$\lbrace\alpha,\beta,\gamma,r_s/\lambda_p,\rho_s\rbrace$ for each
component are:

\begin{itemize}

\item 
Dark matter for N-body only:
$\lbrace0.866,3.18,1.18,1.39\times10^{-2},9.19\times10^3\rbrace$ 
\item 
Dark matter for N-body/Gas:  
$\lbrace0.141,2.78,1.95,2.82\times10^{-2},6.15\times10^2\rbrace$
\item 
Gas:
$\lbrace3.07\times10^{-3},2.84,1.98,2.47\times10^{-2},8.06\times10^2\rbrace$

\end{itemize} 

The density profiles shown in the right panel of Figure 2 show that when
gas is added to the simulation, the dark matter density profile steepens
at the center. The dark matter parameter $\gamma$, the negative of the
asymptotic
logarithmic slope for inner radii, is $1.18$ for the case with no gas and
$1.95$ for the case with. While the fitting function does not make a
direct comparison of the fits at inner radii easy because of the
difference in scale radius $r_s$ for each component, it does show that the
halo in the case with gas has a steeper inner slope.  As such, this
effect can only worsen the disagreement between the predicted singular
profiles and observations of constant density cores. More detailed
analysis at higher resolution is needed before this conclusion can be made
with certainty, since the part of the profile in question is at radii very
close to the numerical softening length for the gravity calculation, even
though there were more than $10^5$ particles each of gas and dark matter
within $r_{200}$ at this epoch.

\section{Summary and Discussion}
\label{sec:summary} 

We have analyzed the structure of the halo which forms by gravitational
instability when a cosmological pancake is perturbed during its collapse,
based upon ASPH/P$^3$M simulations.  Such halos resemble those formed in
CDM
simulations, with a density profile like that of NFW, when no gas is
included.  Our results strengthen the conclusion reached elsewhere that
halos with a universal structure like that identified by NFW arise from
gravitational instability under a wider range of circumstances than those
involving hierarchical clustering \cite{Huss99,Moore99,Tit99}.
When gas is included, we find that the halo gas acquires a steeper inner
profile than that of the pure N-body results, almost as steep as
$\rho\ {\propto}\ r^{-2}$, as does the dark matter in this case.  The halo
appears to be in a quasi-isothermal equilibrium state, although matter
continues to fall into the halo, mainly along the filaments.  The
persistence of filamentary substructure in the halo seems to manifest
itself by sustaining the radial motion of the infalling particles even
after ``virialization", in contrast to results found from simulations of
hierarchical clustering \cite{ENF98}.  The inclusion of gas lessens this
effect somewhat at the center of the halo.

Our result that the presence of a gasdynamical component steepens the
inner profile of the dark matter halo relative to the pure N-body result
is consistent with some other recent simulations of cluster formation in
the CDM model (e.g Lewis {\em{et al.}} 1999).  Tittley and Couchman (1999)
report $\rho_{DM}\ {\propto}\ r^{-1.8}$ at small r in their simulations,
in
rough agreement with our result, although they note that the NFW profile
is also a reasonable fit. However, for CDM-like initial 
conditions, they actually report a steeper inner profile for the gas than
for the dark matter, while our gas and dark matter inner profiles are
equally steep. CDM simulations
of a single cluster by many different codes, summarized in Frenk {\em{et
al.}} (2000), report somewhat different results from those of Tittley and
Couchman (1999). In particular, while
the dark matter profile was found in Frenk {\em{et al.}} (2000) to be
well-fit by the NFW profile when
gas was included, the gas density profile was even flatter than this
at small radii.  Lewis {\em{et al.}} (1999) also find a gas density
inner profile which is flatter than that of the dark matter.  It
would be tempting to attribute the difference
between our result that the density profiles of both gas and dark matter 
are steeper than NFW when gas is included, and the flatter inner profiles
reported in Frenk {\em{et al.}}
(2000), to the difference between CDM-like and pancake-instability initial
conditions. However, this would not explain the difference between the
results of Frenk {\em{et al.}} (2000) and Tittley and Couchman (1999),
both
for
CDM initial conditions.

\acknowledgements This work was supported in part by grants NASA NAGC-7363
and NAG5-7821, Texas ARP No. 3658-0624-1999, and NSF SBR-9873326 and
ACR-9982297.

\end{document}